\documentclass[prb,amsmath,amssymb,twocolumn,showpacs,superscriptaddress,longbibliography]{revtex4-1}
\usepackage{amsfonts,amsmath,amssymb,bm}
\usepackage{graphicx,graphics,float}
\usepackage{bm}
\usepackage{amssymb}
\usepackage{colordvi}
\usepackage{graphicx}
\usepackage{color}
\usepackage[colorlinks=true,linkcolor=blue,citecolor=blue,urlcolor=blue]{hyperref}%
\usepackage{hyperref}
\usepackage{comment}
\usepackage{harpoon}

\begin{document}

\title{Unconventional four-terminal thermoelectric transport due to inelastic transport: cooling by transverse current, transverse thermoelectric effect and Maxwell demon}

\author{Jian-Hua Jiang}\email{jianhuajiang@suda.edu.cn}
\affiliation{School of physical science and technology \&
Collaborative Innovation Center of Suzhou Nano Science and Technology, Soochow University, Suzhou 215006, China.}

\author{Jincheng Lu}\email{jincheng_lu@qq.com}
\affiliation{School of physical science and technology \&
Collaborative Innovation Center of Suzhou Nano Science and Technology, Soochow University, Suzhou 215006, China.}

\author{Yoseph Imry}\email{Deceased.}
\affiliation{Department of Condensed Matter Physics, Weizmann Institute of Science, Rehovot 76100, Israel}

\date{\today}
\begin{abstract}
We show that in mesoscopic four-terminal thermoelectric devices with two electrodes (the source and the drain) and two heat baths, inelastic scattering processes can lead to unconventional thermoelectric transport. The source (or the drain) can be cooled by passing a thermal current between the two heat baths, with no net heat exchange between the heat baths and the electrodes. This effect, termed as cooling by heat current, is a mesoscopic heat drag effect. In addition, there is a transverse thermoelectric effect where electrical current and power can be generated by a transverse temperature bias (i.e., the temperature bias between the two heat baths). This transverse thermoelectric effect, originates from inelastic scattering processes, may have advantages for improved figures of merit and power factor due to spatial separation of charge and heat transport. We study the Onsager current-affinity relations, the linear-response transport properties, and the transverse thermoelectric figure of merit of the four-terminal thermoelectric devices for various system parameters. In addition, we investigate the efficiency and power of the cooling by transverse current effect in both linear and nonlinear transport regimes. We also demonstrate that by exploiting the inelastic transport in the quantum-dot four-terminal systems, a type of Maxwell's demon can be realized using nonequilibrium heat baths.
\end{abstract}


\maketitle

\section{Introduction}

The study of thermoelectric transport at nanoscales\cite{DubiRMP,Nanotechnology,JiangCRP,thierCRP,BENENTI20171} is important at least for two reasons. First, it is a realm where mesoscopic fluctuations and dissipations work together with quantum mechanics\cite{natureVenka,DavidPRL,Schaller2013,Jordan2013,JiangPRX,SanchezPRL,BijayJiang,Thier2015,JiangBijayPRB17,Mu2017,CooperativeSpin,jaliel-exper,CPB}. Second, nanostructured materials are important routes toward high thermoelectric efficiency and power, as motivated by the seminal works of Hicks and Dresselhaus\cite{HicksPRB1,HicksPRB2,VenPRB,Biswas2012}. While most of the studies are based on elastic transport processes, recent researches found that inelastic transport processes lead to phenomena that have not been found in elastic transport\cite{OraPRB2010,David2011PRB,Lena2012,SothmannPRB,JiangSR,Rongqian,JiangNearfield,CooperativeSpin}. The first example is the cooling by heating effect in three-terminal (i.e., two electronic electrodes and a bosonic terminal) thermoelectric transport where one of the two electronic reservoirs can be cooled (the other one heated) by heat injection from the bosonic reservoir\cite{Cooling1,Cooling2,Cooling3}. The second example is the linear thermal transistor effect, namely that using three-terminal inelastic thermoelectric transport thermal transistor effect can be realized in linear-response regime without relying on nonlinear negative differential thermal conductance\cite{Jiangtransistors,lu-PRB}.

Inelastic thermoelectric transport realizes high-efficiency and powerful thermoelectric devices. As shown in Refs.~[\onlinecite{JiangNJP}] and [\onlinecite{JiangJAP}], high figure of merit inelastic thermoelectric devices requires only small bandwidth of the bosons involved in the inelastic processes. Unlike Mahan and Sofo's proposal of using narrow electronic bandwidth~\cite{Mahan}, such a requirement does not cause suppressed electrical conductivity and power~\cite{ZhouPRL,Jiang2013,JiangOra}, if the interaction between electrons and bosons are strong. Such inelastic thermoelectric devices can have large figures of merit and high power factors. It was found recently by Whitney et {\it{al.}}~\cite{WhitneyPhysE} that in four-terminal thermoelectric transport through a quantum dot (QD) (i.e., the source, the drain, and two electrodes capacitively coupled with the QD as thermal baths), electrical powers can be generated even when the total heat injection from the two thermal baths vanishes. Such nontrivial thermoelectric conversion is realized through Coulomb drag, which is due to inelastic Coulomb scattering processes.

In this work we show that a nontrivial phonon drag effect, termed as ``cooling by transverse current`` can emerge in four-terminal thermoelectric systems with two electrodes and two thermal (phonon) baths (can also be baths with other collective excitations). By ``cooling by transverse current`` effect, we mean that one of the two electronic reservoirs (the source or the drain) can be cooled by passing a heat current between the two thermal baths. The crucial transport processes are the inelastic transitions assisted by the collective excitations in the thermal baths. These inelastic transitions also yield other anomalous transport effects such as transverse thermoelectric effects (i.e., electrical current and power can be induced by a transverse temperature gradient). Here the transverse thermoelectric effect is like the Nernst-Ettingshausen effect, but not due to time-reversal (Onsager) symmetry broken by magnetic field or magnetization, but rather due to inelastic transitions and chirality broken. The ``cooling by transverse current`` is similar to thermal Hall effect, which is again not due to time-reversal (Onsager) symmetry broken, but rather due to the inelastic transport and chirality broken. These unconventional transport effects are promising for heat control at nanoscale as well as high-performance thermoelectric devices. The spatial separation of heat and charge flow, as well as the enriched controllability, enables more degrees of freedom to manipulate thermoelectric transport. In this work, we first present our studies on the linear thermoelectric transport and the thermoelectric figure of merit and power factor (and conditions for high thermoelectric performance) in the four-terminal inelastic thermoelectric devices. We continue on studying the cooling by transverse current effect with emphasizing on the cooling efficiency and power beyond the linear-responses. We then show that the four-terminal inelastic  devices can realize a type of Maxwell's demon using nonequilibrium baths.

\begin{figure}[htb]
\begin{center}
\centering\includegraphics[width=7.0cm]{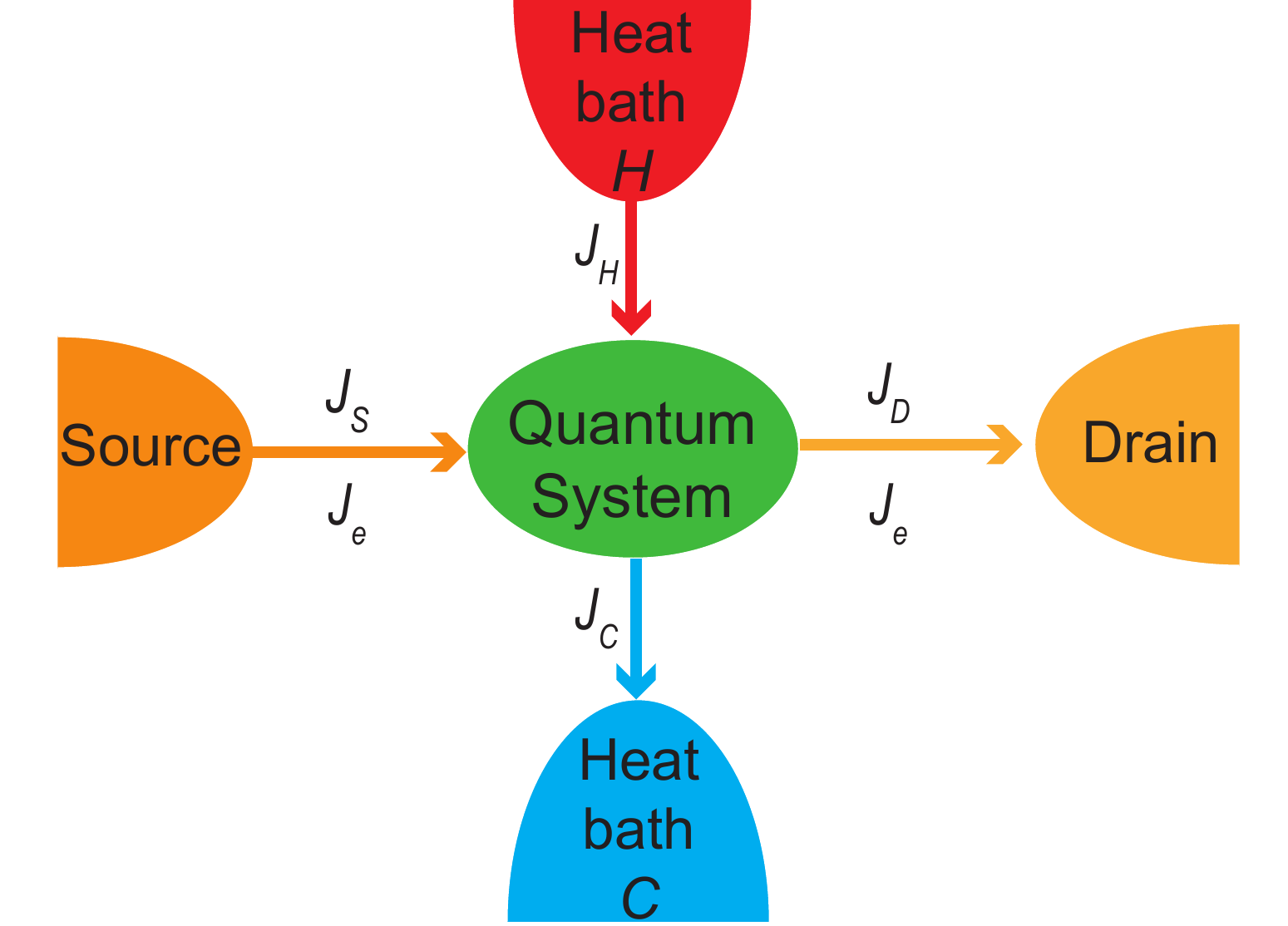}
\caption{Schematic of a four-terminal mesoscopic thermoelectric device. The source and the drain are electronic reservoirs. The other two terminals serve as the
heat baths which only provide collective excitations to the central quantum system (i.e., provide only energy output or input). They do not exchange charge with the central quantum system nor other reservoirs. One heat bath $H$ has higher temperature than the other heat bath $C$. These heat baths can be bosonic (e.g., phonon bath) or electronic (e.g., electronic bath with charge fluctuations as the collective excitations).  Four heat currents $J_S$, $J_H$, $J_C$, $J_D$ and the electric current $J_e$ are illustrated.} ~\label{fig:system}
\end{center}
\end{figure}

\section{currents and affinities}
We consider a four-terminal thermoelectric device as illustrated in Fig.~\ref{fig:system}. The electrical current $J_e$ flows from source to drain. Besides, there are four heat currents, $J_S$, $J_D$, $J_H$, and $J_C$. Only three of them are independent, due to energy conservation, i.e.,
\begin{equation}
J_S + \frac{\mu_S}{e}J_e + J_H = J_C + J_D + \frac{\mu_D}{e}J_e,
\label{eq:EnergyCon}
\end{equation}
where $e$ is the carrier charge, and $\mu_i$ ($i = S,D$) are the electrochemical potentials of the two electrodes. The three independent heat currents are chosen as the heat current flowing out of the source $J_S$, and the symmetric and anti-symmetric combinations of $J_H$ and $J_C$, defined as,
\begin{equation}
J_{in}\equiv J_H-J_C, \quad J_q \equiv \frac{J_H + J_C}{2}.
\end{equation}
$J_{in}$ is regarded as the total heat current injected into the central quantum system from the two thermal baths, while $J_q$ is the heat exchange current between the two heat baths intermediated by the central quantum system.

The affinities can be found via examining the entropy production,
\begin{equation}
\frac{dS}{dt} = -\frac{J_S}{T_S} - \frac{J_H}{T_H} + \frac{J_D}{T_D} + \frac{J_C}{T_C},
\end{equation}
where $T_i$ ($i=S,D,H,C$) are the temperatures of the four reservoirs. Inserting Eq.~\eqref{eq:EnergyCon}, we obtain
\begin{equation}
\begin{aligned}
\frac{dS}{dt} =& J_S\left(\frac{1}{T_D} - \frac{1}{T_S}\right) + J_{in}\left(\frac{1}{T_D} - \frac{1}{2T_H} - \frac{1}{2T_C}\right)\\
& + J_q\left(\frac{1}{T_C} - \frac{1}{T_H}\right) + J_e\frac{\mu_S-\mu_D}{eT_D}\\
&\equiv J_SA_S + J_{in}A_{in} + J_qA_q + J_eA_e.
\end{aligned}
\end{equation}
Hence the affinities are
\begin{equation}
\begin{aligned}
&A_S \equiv \frac{1}{T_D} - \frac{1}{T_S}, \quad A_{in} \equiv \frac{1}{T_D} - \frac{1}{2T_H} - \frac{1}{2T_C}, \\
&A_q \equiv \frac{1}{T_C} - \frac{1}{T_H}, \quad A_e \equiv \frac{\mu_S-\mu_D}{eT_D}.
\end{aligned}
\end{equation}
From the above one can see that the heat current $J_{in}$ is driven by the temperature difference between the drain and the two heat baths. In the linear-response regime, $A_{in}\approx (T_{av}-T_D)/T^2$ where $T_{av}\equiv\frac{1}{2}(T_H + T_C)$ is the average temperature of the two heat baths and $T$ is the equilibrium temperature of the whole system. In comparison, $J_q$ is driven by the temperature difference between the two heat baths, i.e., in linear-response, $A_q\approx (T_H-T_D)/T^2$. The affinities for the charge and heat currents flowing out of the source are known in conventional two-terminal thermoelectric transport.

\begin{figure}[htb]
\begin{center}
\centering\includegraphics[width=7.0cm]{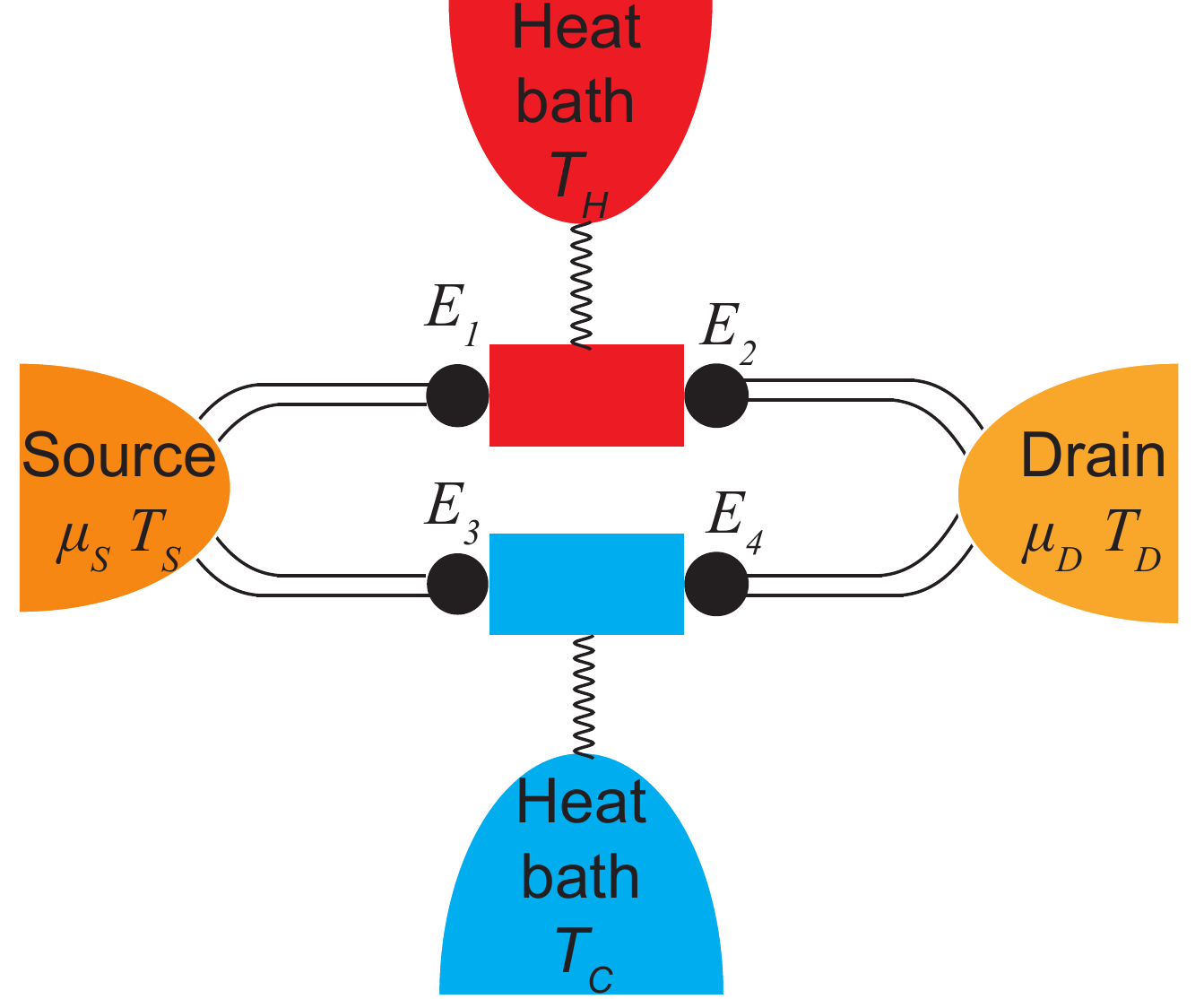}
\caption{Schematic of QDs four-terminal thermoelectric devices. There are two parallel transport channels. Each channel has two QDs with different energies and a heat bath to enable inelastic transport. The two channels are spatially separated so that the heat bath $H$ ($C$) couples only to the upper (lower) channel.} ~\label{fig:4T-model}
\end{center}
\end{figure}

\section{microscopic system and realization}

We consider a simple microscopic system illustrated in Fig.~\ref{fig:4T-model} and show that several unconventional thermoelectric phenomena can emerge. The system consists of four QDs: QDs 1 and 2 (of electronic energy $E_1$ and $E_2$) are coupled with the hot heat bath $H$, while QDs 3 and 4 (of energy $E_3$ and $E_4$) are coupled with the cold heat bath C. We restrict our discussions on the situations where there is only one energy level (two
levels if spin degeneracy is taken into account) in each QD that is relevant for the transport. We also assume that the temperature is high and the Coulomb blockade and other interaction effect is irrelevant, which is different from the situations studied in Ref.~[\onlinecite{WhitneyPhysE}]. The inelastic phonon-assisted transport is dominant in the high-temperature regime considered in this paper. To suppress elastic sequential tunneling, we consider QDs with mismatched energies, i.e., $E_1\ne E_2$ and $E_3\ne E_4$. Beside phonon-assisted transport, there are several other realizations which include inelastic transport assisted by charge fluctuations\cite{Rafael}, photons\cite{Ruokola,lu-PRB}, and magnons\cite{Sothmann_2012,TangPRB}. The relevant interactions in those realizations are Coulomb interactions, electron-photon interactions, and exchange interactions situations. The heat baths can be bosonic (e.g., phonon baths) or electronic (e.g., electrodes capacitively coupled with the central quantum system). In addition to QDs, $p-n$ junctions with very narrow bands (i.e., $\le0.1eV$) can also serve the same purpose of suppressing the elastic transport and enabling the inelastic transport through photon-assisted (or phonon-assisted) interband transitions. In all these realizations, the temperatures of the four reservoirs can be controlled by local heating or cooling. The substrate also affects the local temperatures of these reservoirs. Their temperatures can be determined via noise measurements. The two transport channels, QDs 1 and 2 vs. QDs 3 and 4, are independent of each other.

In this paper we choose to work with phonon-assisted hopping transport between energy-mismatched QDs\cite{Jiang2012,Jiangtransistors}.
The hopping particle currents through the two independent channels are calculated as,
\begin{equation}
\begin{aligned}
I_{12} &\equiv \Gamma_{1\rightarrow2} - \Gamma_{2\rightarrow1}\\
&=\gamma_{eq}[f_1(1-f_2)N_{12}^+ - f_2(1-f_1)N_{12}^-],
\label{eq:I12}
\end{aligned}
\end{equation}
\begin{equation}
\begin{aligned}
I_{34} &\equiv \Gamma_{3\rightarrow4} - \Gamma_{4\rightarrow3}\\
&=\gamma_{eq}[f_3(1-f_4)N_{34}^+ - f_4(1-f_3)N_{34}^-].
\label{eq:I34}
\end{aligned}
\end{equation}
Here $I_{12}$ is the particle current flowing from the QD 1 to the QD 2, $I_{34}$ is the particle current flowing from the QD 3 to the QD 4. $\Gamma_{1\rightarrow2}$ ($\Gamma_{3\rightarrow4}$)  is the electron transfer rate from QD 1 to QD 2 (QD 3 to QD 4). $\gamma_{eq}$ is the phonon-assisted hopping rate calculated from the Fermi golden rule. $N_{12}^{\pm}\equiv|n_B(\pm\omega_u/k_BT_H)|$ and $N_{34}^{\pm}\equiv|n_B(\pm\omega_d/k_BT_C)|$ with $n_B\equiv1/(e^x-1)$ and
\begin{equation}
\omega_u\equiv E_2-E_1, \quad \omega_d \equiv E_4-E_3.
\end{equation}
The steady-state electronic distributions on the QDs, $f_i$ ($i = 1,2,3,4$), can be determined by the master equation approach. We shall rather consider a simpler situation under the following assumptions. We assume that the contacts between the source and the two QDs 1 and 3 can be made very good. Under such conditions, we can approximate the distributions on the two QDs 1 and 3 as $f_1\approx n_F\left(\frac{E_1-\mu_S}{k_BT_S}\right)$ and $f_3\approx n_F\left(\frac{E_3-\mu_S}{k_BT_S}\right)$, where $n_F\equiv1/(e^x+1)$  is the Fermi-Dirac distribution. Similarly, if the contacts between the drain and the two QDs 2 and 4 can be made very good, the distributions on the QDs 2 and 4 can be approximated as $f_3\approx n_F\left(\frac{E_2-\mu_D}{k_BT_D}\right)$ and $f_4\approx n_F\left(\frac{E_4-\mu_D}{k_BT_D}\right)$.

When the linewidth of the QDs are much smaller than $k_BT$, we can ignore the linewidth of the QDs. In this regime, there are relationships between the currents,
\begin{equation}
J_S = E_1I_{12} + E_3I_{34}, \quad J_H = \omega_uI_{12}, \quad J_C = -\omega_dI_{34}.
\end{equation}
We then have
\begin{equation}
J_{in} = \omega_uI_{12} + \omega_dI_{34}, \quad J_q = \frac{1}{2}(\omega_uI_{12} - \omega_dI_{34}).
\end{equation}
Linear expansion of the currents in Eqs.~\eqref{eq:I12} and \eqref{eq:I34} in terms of the affinities, and using the above equations, we obtain the following linear-response transport equations
\begin{equation}
\begin{aligned}
&J_i = \sum_jM_{ij}A_j, \quad i,j=e,S,in,q\\
&M_{ij} = \frac{\gamma_0}{k_B}\langle a_ia_j \rangle,\\
&a_e = e, \, a_S = E_l,\, a_{in} = \omega, \, a_q = \frac{1}{2}\omega\sigma.
\label{eq:JMij}
\end{aligned}
\end{equation}
Here $\Gamma_0\equiv\Gamma_{1\rightarrow2}^0 + \Gamma_{3\rightarrow4}^0$ where the superscript 0 denotes the quantities calculated at thermal equilibrium. For instance, the electrical conductance is given by $G=M_{e,e}T$ and (one of) the heat conductance is $K_S=M_{S,S}T$. The average in the above equation is performed between the two channels
\begin{equation}
\langle...\rangle = \sum_{j=u,d}p_j...
\end{equation}
with $p_u\equiv\Gamma_{12}^0/\Gamma^0$ for the up channel and $p_d\equiv\Gamma_{34}^0/\Gamma^0$ for the down channel and $p_u+p_d=1$. Therefore, the
weights for the average are proportional to the hopping rate (conductance) in each channel. $E_l$ in Eq.~\eqref{eq:JMij} denotes the energy of the QD on the left side of each channel, i.e., $E_{lu} = E_1$ for the up channel and $E_{ld} = E_3$ for the down channel. The integer index $\sigma$ is used to label the two channels: $\sigma=1$ for the up channel and $\sigma=-1$ for the down channel.

\section{conventional and unconventional thermoelectric energy conversions}
In the four-terminal thermoelectric device there are three Seebeck coefficients
\begin{equation}
\begin{aligned}
S_S &= \frac{k_B}{e}\frac{p_uE_1+p_dE_3}{k_BT},\\
S_{in} &= \frac{k_B}{e}\frac{p_u\omega_u+p_d\omega_d}{k_BT},\\
S_q &= \frac{k_B}{e}\frac{p_u\omega_u - p_d\omega_d}{2k_BT},
\end{aligned}
\end{equation}
representing electrical current (power) generation induced by three temperature gradients which originate from the three thermal affinities, $A_s$, $A_{in}$ , and $A_q$. An important question is which symmetry needs to be broken to get finite Seebeck coefficients. For example, a finite S S needs the breaking of ``particle-hole`` symmetry, ${\mathcal{T}}_{PH}$, around the electrochemical potential\cite{Jiangtransistors}. A finite $S_{in}$ is related to broken left-right mirror symmetry, ${\mathcal{T}}_{Mx}$\cite{Jiangtransistors}. From the above equations, both the left-right ${\mathcal{T}}_{Mx}$ and
up-down ${\mathcal{T}}_{My}$ mirror symmetries need to be broken to
have finite $S_q$. Interestingly, $S_q$ is invariant under the
multiplicative operator ${\mathcal{T}}_I={\mathcal{T}}_{Mx}*{\mathcal{T}}_{My}$ (we term ${\mathcal{T}}_I$ as inversion operator/symmetry), but $S_q$ changes sign under each of the mirror operator. In geometry, the broken of the two mirror symmetries while preserving the inversion symmetry is associated with chirality broken.  Therefore, one can conclude that the Seebeck coefficient $S_q$ is associated with chirality broken.

We term $S_q$ as the four-terminal transverse Seebeck coefficient, which describes electrical current $J_e$ generation induced by the perpendicular temperature gradient $T_H-T_C$. This effect, similar to the Nernst-Ettingshausen effect, however, originates not from time-reversal (Onsager) symmetry broken by magnetic field or magnetic interaction, but due to chirality broken. In this transverse thermoelectric effect, the spatial separation of the charge and heat currents is advantageous since electrical and thermal conduction can be manipulated independently. High figure of merit can be attained by suppressing the heat conductivity while enhancing the electrical conductivity and Seebeck coefficient\cite{,JiangNJP,JiangJAP}. Explicitly, the figure of merit for this transverse thermoelectric effect is
\begin{equation}
Z_qT = \frac{M_{e,q}^2}{M_{e,e}M_{q,q}-M_{e,q}^2}.
\label{eq:ZqT}
\end{equation}
The power factor is defined as $P_q=GS_q^2$. From the above we can see that large figures of merit can be achieved when the following two types of unwanted heat/charge conduction are small: (1) the heat conduction from $H$ to $C$ that does not contribute to thermoelectric conversion, (2) the charge conduction between the source and the drain due to elastic processes which has $a_q = 0$. The former phonon heat conduction can be reduced by, e.g., enhancing phonon scattering with random interfaces and disorders between the $H$ and $C$ heat baths (or using materials with low phonon heat conductivity). The latter charge conduction leads to Joule heating. This charge
conduction due to elastic processes can be suppressed by increase the energy difference between the two QDs in each channel, or by enhancing the inelastic processes via strong electron-phonon interaction (such as using GaN or GaAs QDs where electron-phonon coupling is strong) or high temperature\cite{Jiang2012}. It has been shown in Refs.~\cite{JiangJAP} and \cite{JiangPRL}, a proper energy difference $E_2 - E_1$ can effectively suppress the elastic conduction. In general, Eq.~\eqref{eq:ZqT} predicts high thermoelectric figure-of-merit if the variance of the energy $a_q$ is small, while its average is large. In the limit of zero variance, infinity figure-of-merit gives the Carnot efficiency, which denotes the limit without dissipation\cite{JiangOra}.

A configuration favoring the figure-of-merit and power factor for the transverse thermoelectric effect is to have $\omega_u=-\omega_d$, i.e., opposite energy difference for the up and down channels. We also notice that a configuration with ``quadrapole-type`` symmetry, i.e., $E_1-E_2=-E_3=E_4$ can lead to $S_S=S_{in}=0$, but $S_q\ne0$. In comparison, a ``dipole-type`` configuration $E_1-E_2=E_3=-E_4$ favors thermoelectric conversion through $S_{in}$, as has been shown in Ref.~\cite{JiangJAP}. To make the symmetry more explicit, we write the figure of merit for the three thermoelectric effects in terms of average over the energy in each channels,
\begin{subequations}
\begin{align}
Z_ST &= \frac{{\langle a_S \rangle}^2}{{\langle a^2_S \rangle}-{\langle a_S \rangle}^2}, \quad a_S=E_l,\\
Z_qT &= \frac{{\langle a_q \rangle}^2}{{\langle a^2_q \rangle}-{\langle a_q \rangle}^2}, \quad a_q = \frac{1}{2}\omega\sigma, \\
Z_{in}T &= \frac{{\langle a_{in} \rangle}^2}{{\langle a^2_{in} \rangle}-{\langle a_{in} \rangle}^2}, \quad a_{in}=\omega.
\end{align}
\end{subequations}

\section{cooling by transverse current}
We discover a mode of cooling in our system, i.e., cooling, say, the source by driving a heat current between the heat baths $H$ and $C$ (shall be termed as ``cooling by transverse current``. This is different from the previous ``cooling by heating`` effect where cooling is driven by a finite heat current injected into the quantum system. In the cooling by transverse current effect, heat injected into the quantum system is not necessary, since the driving force of the cooling is the energy exchange between the two heat baths via the central quantum system.

It is convenient to demonstrate the cooling by transverse current effect in the situations with
\begin{equation}
A_e = A_{in} = 0.
\label{eq:Ain0}
\end{equation}
The two heat currents of concern are then given by~\cite{Jiang2013,JiangPRL}
\begin{subequations}
\begin{align}
J_S = M_{S,S}A_S + M_{S,q}A_q,\\
J_q = M_{S,q}A_S + M_{q,q}A_q.
\end{align}
\end{subequations}
cooling by transverse current is to cool the source (i.e., $J_S>0$ despite $A_S<0$) by a positive $J_q$ (we assume $T_H>T_C$, i.e., $A_q>0$). To have $J_S>0$ while $A_S<0$, one needs
\begin{equation}
M_{S,q}>0.
\end{equation}
We propose to fulfill such a requirement via the following energy configuration,
\begin{equation}
E_1=\varepsilon, \quad E_2 = \varepsilon+\omega, \quad E_3 = \varepsilon+\omega, \quad E_4=\varepsilon,
\end{equation}
for $\varepsilon\omega>0$. We shall consider the case with both $\varepsilon>0$ and $\omega>0$. The situation with both $\varepsilon$ and $\omega$ negative gives the same results. There are lots of other energy configurations that can realize cooling by transverse current effect. Here we consider the above energy configuration to simplify our discussions.

The phenomenon that the heat flow from $H$ to $C$, $J_q$, is able to drive a cooling of the source is closely related to the inelastic thermoelectric transport processes in our system. It is very similar to a phonon drag effect (we may also call it as ``cooling by transverse current`` effect). This effect is also a transverse thermal conduction effect, i.e., heat current is induced in the direction perpendicular to the temperature gradient. We remark that the transverse thermal conduction includes the situations with $M_{S,q} = 0$ which do not have the cooling (of the source) by thermal current effect. We remark that the conditions for cooling of the drain are different from the conditions for cooling the source.

We shall henceforth focus on situations restricted by Eq.~\eqref{eq:Ain0}. The restriction $A_{in} = 0$ establishes a relation between $T_H$ and $T_C$, so only one of them is independent. We may regard $T_H$ as an independent variable, whereas $T_C$ is determined by
\begin{equation}
T_C=\left(\frac{2}{T_D}-\frac{1}{T_H}\right)^{-1}.
\end{equation}
We set $T_S<T_D$ and aim at cooling the source (i.e., $J_S>0$) using $J_q$ through the mesoscopic heat thermal current effect.

The coefficient of performance (COP) for the cooling by transverse current effect in our four-terminal system is given by\cite{MyJAP,trade-off}
\begin{equation}
\eta_{COP}=\frac{J_S}{J_q}.
\end{equation}
Microscopically, the currents are
\begin{equation}
J_S = \varepsilon I_{12}+(\varepsilon+\omega)I_{34}, \quad J_q = \frac{1}{2}\omega(I_{12}+I_{34}).
\end{equation}
In the strong-coupling limit (SCL)\cite{Kedem,Cooling2}, $I_{12}=I_{34}$ (i.e., $J_{in}=0$\cite{WhitneyPhysE}), the COP is
\begin{equation}
\eta^{SCL}_{COP} = \frac{2\varepsilon+\omega}{2\omega}.
\end{equation}
The entropy production of the system is\cite{JiangPRL,Jiang2017}
\begin{equation}
\frac{dS}{dt} = J_SA_S + J_qA_q,
\end{equation}
since $A_{in}=0$ and $A_e=0$. Therefore, we have
\begin{equation}
\eta_{COP} = \left(T_SA_q - \frac{T_S}{J_q}\frac{dS}{dt}\right)\frac{T_D}{T_D-T_S}.
\end{equation}
The reversible COP is
\begin{equation}
\eta^{rev}_{COP} = \left(\frac{T_S}{T_C}-\frac{T_S}{T_H}\right)\frac{T_D}{T_D-T_S}=-\frac{A_q}{A_S}.
\end{equation}
The above relationship clearly demonstrates that the cooling of the source is driven by $A_q$. In other words, the negative entropy production associated with the cooling of the source, $J_SA_S<0$, is compensated by the positive entropy production of $J_qA_q$, in agreement with Kedem and Caplan\cite{Kedem}.

\begin{figure}[htb]
\begin{center}
\centering \includegraphics[width=8.5cm]{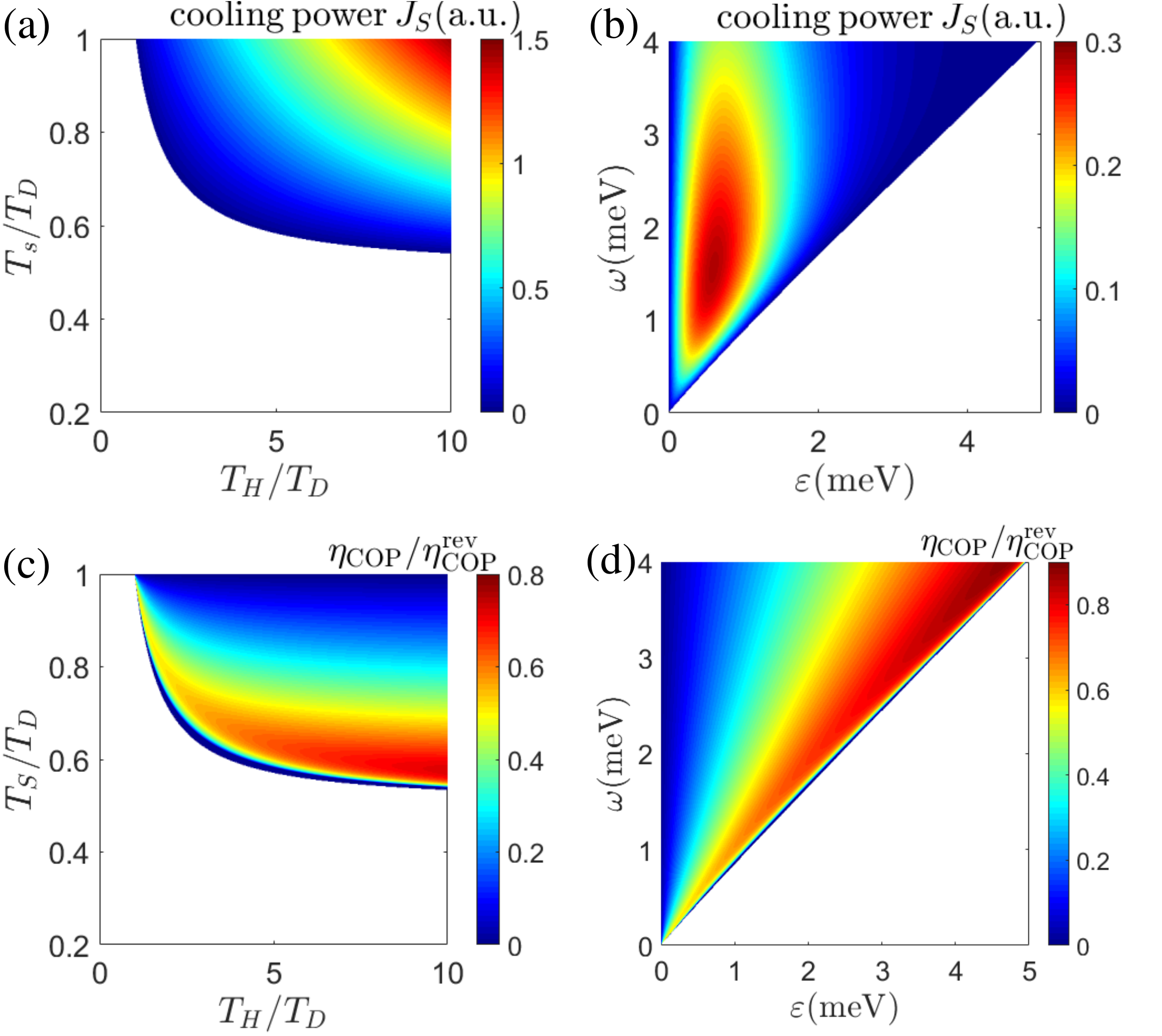}
\caption{(a) Cooling power $J_S$ as functions of temperature ratios $T_S/T_D$ and $T_H/T_D$ for $\varepsilon=1\,\rm{meV}$ and $\omega=2\,\rm{meV}$. (b) Cooling power $J_S$ as functions of QD energies $\varepsilon$ and $\omega$ for $k_BT_H=6\,\rm{meV}$ and $k_BT_S=0.6\,\rm{meV}$. (c) COP $\eta_{COP}$ measured in unit of the reversible COP $\eta_{COP}^{rev}$ for the cooling by transverse current effect as a function of the temperature ratio $T_S/T_D$ and $T_H/T_D$ when QDs energy is the same as in (a). (d) $\eta_{COP}^{rev}$ as functions of QD energies $\varepsilon$ and $\omega$ for the same parameters as in (b). Common parameters: $\mu=0$, $k_BT_D=1\,\rm{meV}$ and $T_C=1/(2/T_D-1/T_H)$. Throughout this paper, ``a.u.'' denotes ``arbitrary units''.}
~\label{fig:PE-TH}
\end{center}
\end{figure}

It is expected that at reversible COP $\eta^{rev}_{COP}$ the cooling power vanishes, since entropy and the currents are zero\cite{JiangOra,JiangPRE}. In other situations $0<\eta_{COP}<\eta^{rev}_{COP}$. We will regard the temperature of the drain $T_D$ as fixed in the discussions below. In realistic situations $T_D$ can be set by the temperature of the substrate.

The working condition of the cooling by transverse current effect is $J_S>0$ for $A_S<0$. For each given energy configuration, this imposes restrictions on the temperatures $T_H$ and $T_S$. In fact, there is a minimum temperature that we can cool the source down for given $T_H$ and $T_S$. On the other hand, for a given temperature of the source $T_S$, there is a minimum $T_H$ to fulfill the cooling by transverse current effect.

\begin{figure}[htb]
\begin{center}
\centering \includegraphics[width=8.5cm]{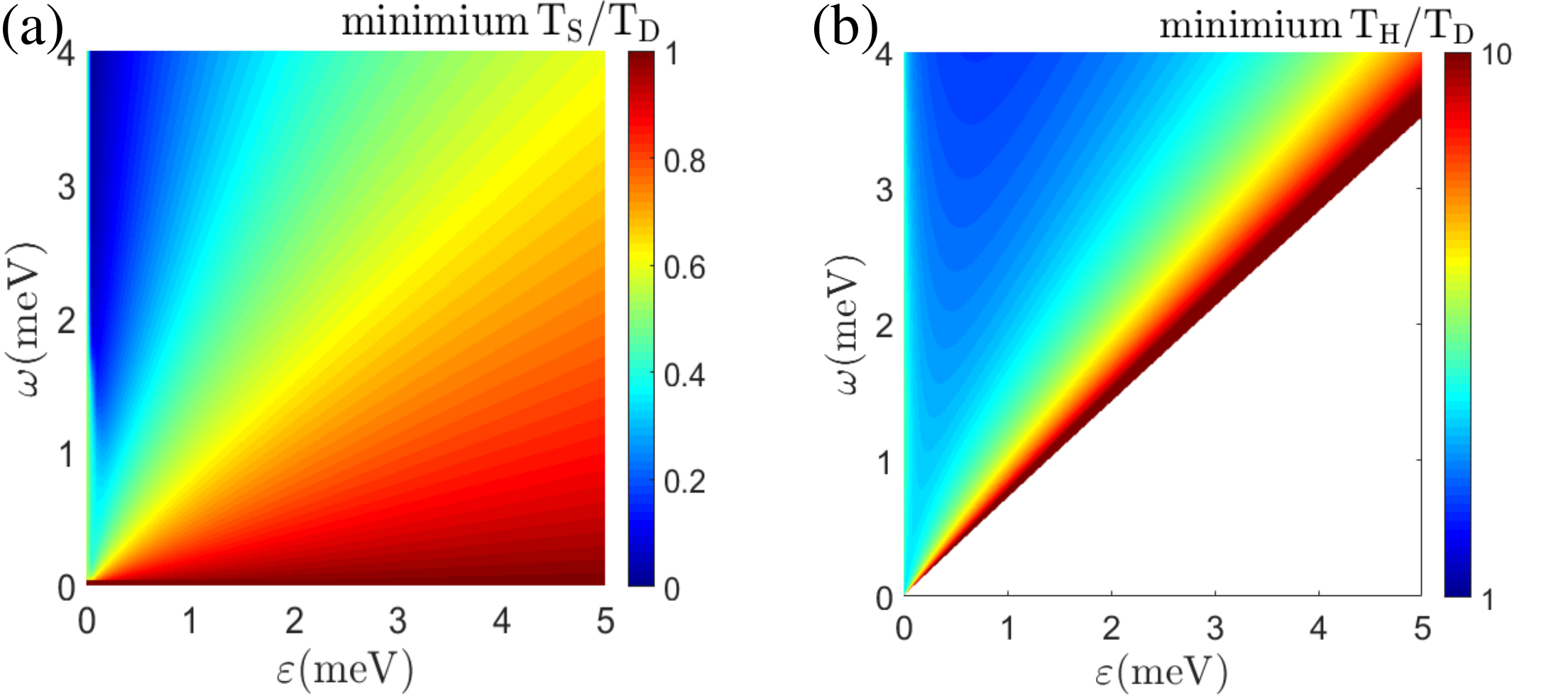}
\caption{(a) The lowest temperature of source $T_S$ that can be cooled down via the cooling by transverse current effect as functions of $\varepsilon$ and $\omega$ for $k_BT_H=6\,\rm{meV}$. (b) The lowest temperature of the hot heat bath $T_H$ that can perform the cooling by transverse current effect as functions of $\varepsilon$ and $\omega$ for $k_BT_S=0.6\,\rm{meV}$ and $\omega\ge0.9\varepsilon$. Other parameters: $\mu=0$, $k_BT_D=1\,\rm{meV}$ and $T_C=1/(2/T_D-1/T_H)$.}
\label{fig:THTS}
\end{center}
\end{figure}

In Fig.~\ref{fig:PE-TH}(a) we plot the cooling power $J_S$ as functions of the temperatures $T_H$ and $T_S$ measured in units of $T_D$. For a given set of $\varepsilon$ and $\omega$, the cooling power increases with both $T_H$ and $T_S$. The condition, $J_S=0$, defines the lowest temperature of the source $T_S$ that can be cooled down for a given $T_H$, or the lowest temperature of the heat bath $T_H$ that starts to cool the source via the thermal current effect at a given source temperature $T_S$. We use the Fermi golden rule, Eqs.~\eqref{eq:I12} and \eqref{eq:I34}, to calculate the currents, power, and efficiency throughout this work.

In Fig.~\ref{fig:PE-TH}(b) we study the cooling power $J_S$ as a function of $\varepsilon$ and $\omega$ for fixed $T_S$ and $T_H$. The results provide useful information in the search of large cooling powers by tuning system parameters. It is found that high cooling powers can be achieved in the region with $\omega\gg\varepsilon$, especially for $\omega<k_BT_H$.

In Fig.~\ref{fig:PE-TH}(c) we show the ratio of the COP over the reversible COP $\eta_{COP}/\eta^{rev}_{COP}$ for a given energy configuration [$\varepsilon=1\,\rm{meV}$ and $\omega=2\,\rm{meV}$, identical to Fig.~\ref{fig:PE-TH}(a)] as a function of the temperature ratios $T_S$ and $T_H$. It is found that high ratios of $\eta_{COP}/\eta^{rev}_{COP}$ appear for low $T_S$ for a given $T_H$. In those regions, the cooling power is very small. In fact, the cooling power is large when $T_S$ is high, i.e., the opposite trend of the COP. Such power-efficiency trade-off is consistent with existing knowledge (particularly as illustrated in Ref.~\cite{JiangPRE} and later confirmed in Refs.~\cite{PED} and \cite{RazPRL}). The optimal ratio $\eta_{COP}/\eta^{rev}_{COP}$ emerges close to the minimum $T_S$ for the cooling by transverse current effect.

The dependence of the ratio $\eta_{COP}/\eta^{rev}_{COP}$ on the QDs energy $\omega$ and $\varepsilon$ [for the same parameters as in Fig.~\ref{fig:PE-TH}(b)] is shown in Fig.~\ref{fig:PE-TH}(d). Large ratios of $\eta_{COP}/\eta^{rev}_{COP}$ appear close to the minimum of $\omega$ for each given $\varepsilon$. Apparently the efficiency has an opposite trend in the dependence of QDs energies. That is, the cooling power $J_S$ is large
for $\omega\gg\varepsilon$, while the COP ratio $\eta_{COP}/\eta^{rev}_{COP}$ is large for $\omega\approx\varepsilon$.

We then study the optimal energy configuration, $\omega$ and $\varepsilon$, that achieves the lowest temperature of the source $T_S$ after sufficiently long time cooling. This is determined by the lowest temperature of the source $T_S$ that gives $J_S = 0$. The results are shown in Fig.~\ref{fig:THTS}(a). Again, the lowest $T_S$ is achieved when $\omega\gg\varepsilon$. In the opposite limit, $\varepsilon\gg\omega$, the cooling by thermal effect is quite ineffective. We also calculate the minimum temperature of the hot heat bath $T_H$ required by the cooling by transverse current effect, i.e., $J_S = 0$, for various energy configurations, as presented in Fig.~\ref{fig:THTS}(b). We find that the region with $\omega\gg\varepsilon$ does not require too high temperature of the heat bath $T_H$ to perform the cooling by heat current. Therefore, a favorable parameter regime for the cooling by transverse current effect is $\omega\gg\varepsilon$ with $\omega\lesssim k_BT_H$.

\begin{figure}[htb]
\begin{center}
\centering \includegraphics[width=8.5cm]{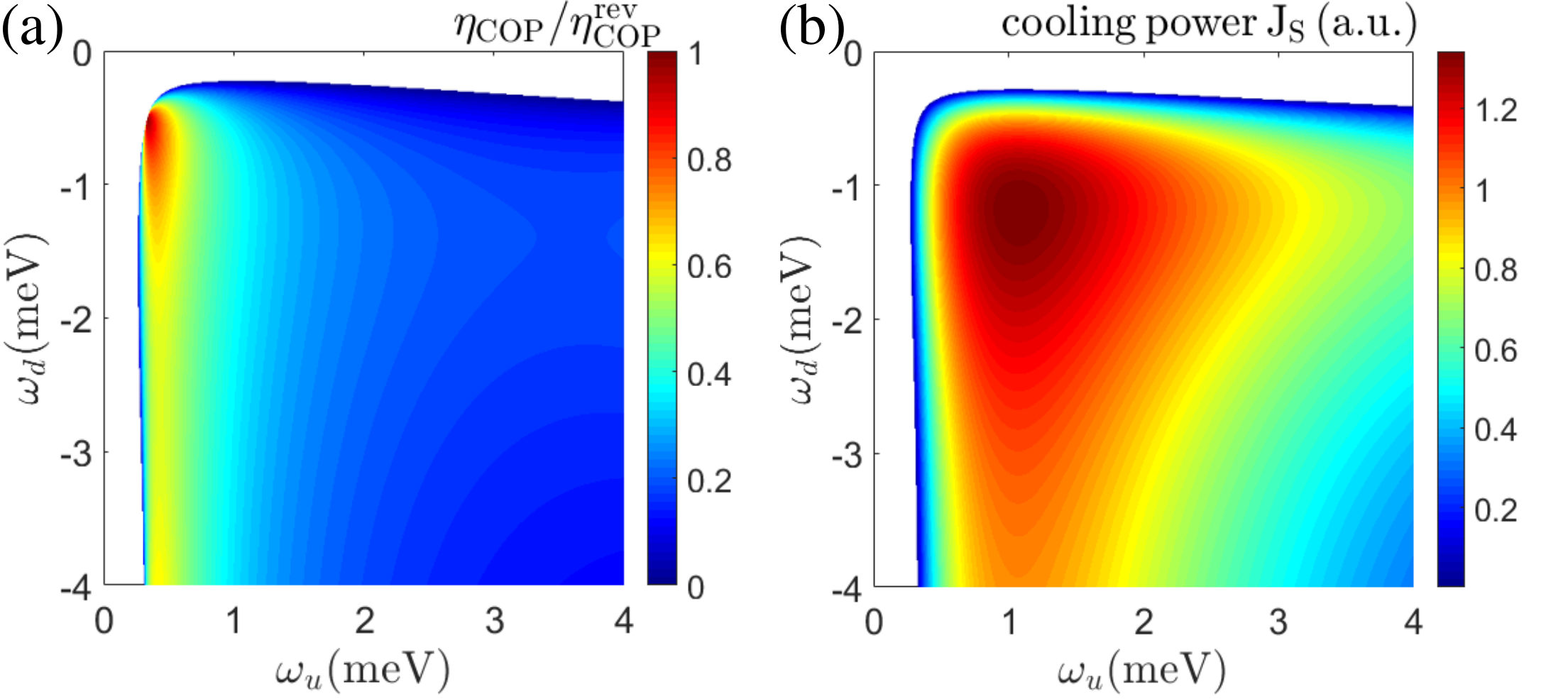}
\caption{(a) COP and (b) cooling power of the cooling by transverse current effect as functions the two energies $\omega_u$ and $\omega_d$. Other parameters: $\mu=0$, $k_BT_D=1\,\rm{meV}$, $k_BT_H=1.5\,\rm{meV}$, $E_1=E_4=1\,\rm{meV}$ and $T_C=1/(2/T_D-1/T_H)$.}
\label{fig:PE-ud}
\end{center}
\end{figure}

In realistic situations the two energies $\omega_u$ and $\omega_d$ may not be equal. We show how the cooling power and COP vary with the two energies, $\omega_u$ and $\omega_d$ in Fig.~\ref{fig:PE-ud}. Both the COP and the cooling power favor the situations with $-\omega_u>\omega_d$. For such a regime, cooling induced by the cold terminal $C$ is more effective, since each phonon emission process gives more energy to the heat bath $C$. The entire picture is that the current in the upper circuit $I_{12}$ is sensitive to phonon temperature $T_H$ and energy $\omega_u$ since it is dominated by the thermal emission of phonons from the heat bath $H$. However, for the current in the lower circuit $I_{34}$, it is mostly determined by the temperature of the source and the drain, while relatively insensitive to the phonon temperature $T_C$ and energy $\omega_d$.

\begin{figure}[htb]
\begin{center}
\centering\includegraphics[width=7.5cm]{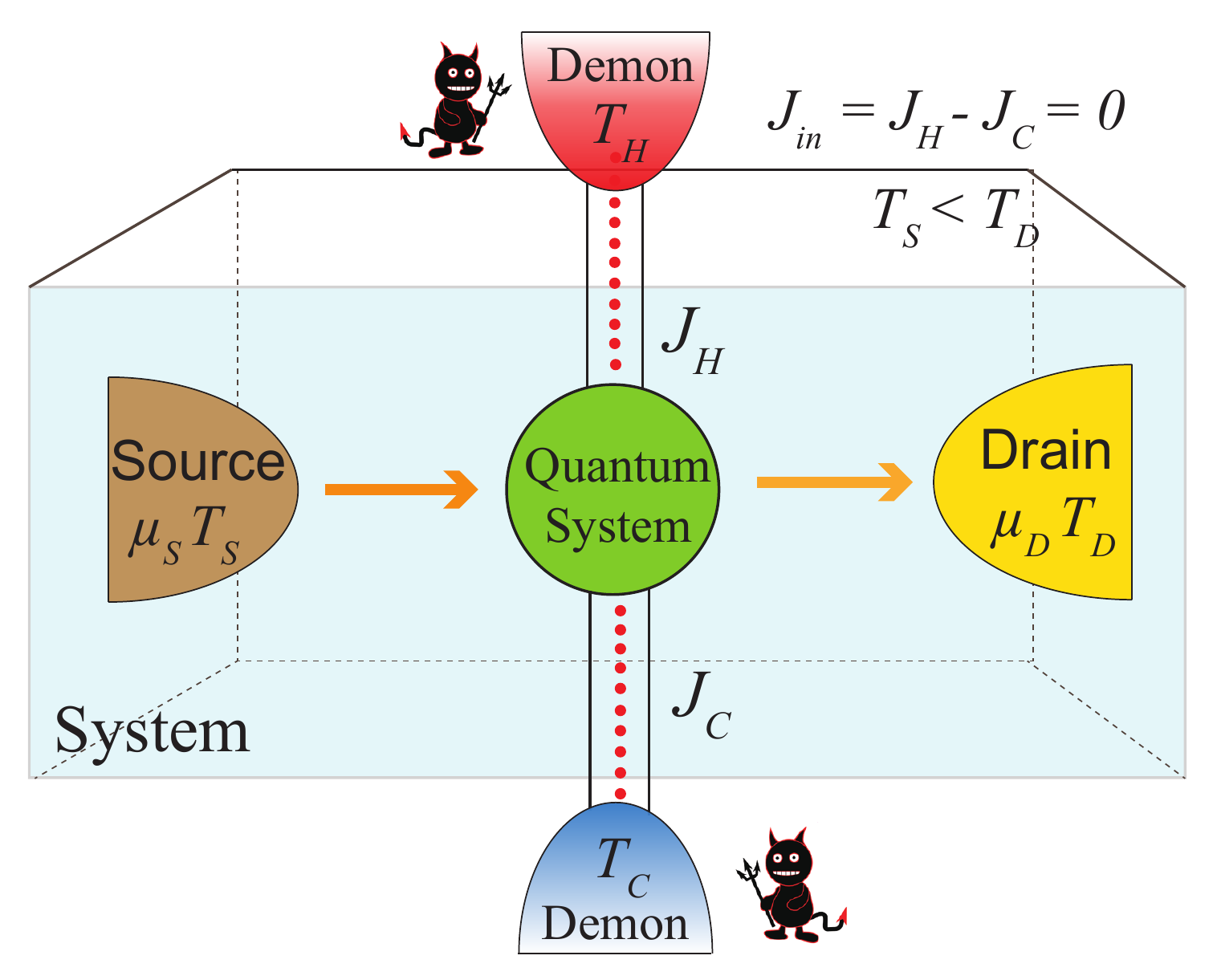}
\caption{Schematic of a four-terminal mesoscopic thermoelectric device as a Maxwell demon. The demon supplies no work or heat to the system, i.e., the total heat current injected into the central quantum system from the two thermal baths is zero, $J_{in}=J_H-J_C=0$.} ~\label{fig:demon-system}
\end{center}
\end{figure}

\section{Four-terminal systems as a Maxwell demon}

The second law of thermodynamics is a general law for macroscopic system which states that the entropy production rate of any macroscopic system cannot be negative. An imagined system with feedback control that violates the conventional formulation of the second law was created by James C. Maxwell~\cite{RMP-demon}, which proposed that a creature (called as ``the Maxwell demon'') with the ability of tracking the velocity of individual gas particles could create a temperature gradient in two macroscopic chambers of gases which are originally at equilibrium with each other. The Maxwell demon can acquire and store the information of the particles. Removing such information, however, as revealed by Landauer~\cite{Landauer}, necessarily yields positive entropy production to restore the second law of thermodynamics. There have been a number of proposals of the Maxwell demon in different forms in various systems~\cite{optical-demon,demon1,demon3,demon2,SanchezPRRes,SanchezDemon}.

In this section, we study a possible implementation of the Maxwell demon in the quantum-dot architecture acting on a system without changing its number of particles or its energy, following the recent works~\cite{SanchezPRRes,SanchezDemon} but in different set-ups and mechanisms. In the four-terminal set-up illustrated in Fig.~\ref{fig:demon-system}, our target is to induce a Maxwell demon based on two nonequilibrium baths (the cold and the hot baths) which can reduce the entropy of the system, the source and the drain, without giving energy or changing the particle number of the system. More explicitly, we aim to induce a heat flow from the cold reservoir, i.e., the source, to the hot reservoir, i.e., the drain, without exchanging energy or particle number with the system (i.e., the source plus the drain). This aim can be achieved by the cooling by transverse current effect at $T_S<T_D$. The entropy reduction in the source and drain is compensated by the larger entropy increase in the cold and hot heat baths. Therefore, the entropy increase rate of the whole system is not negative and the second-law of thermodynamics is not violated.

The condition at which the Maxwell demon neither injects nor extracts heat or energy is
\begin{equation}
J_{in} \,=\, 0.
\end{equation}
Intuitively, the power of the Maxwell demon vanishes when the temperature gradient between the two heat baths $H$ and $C$ becomes zero.

\begin{figure}[htb]
\begin{center}
\centering \includegraphics[width=8.5cm]{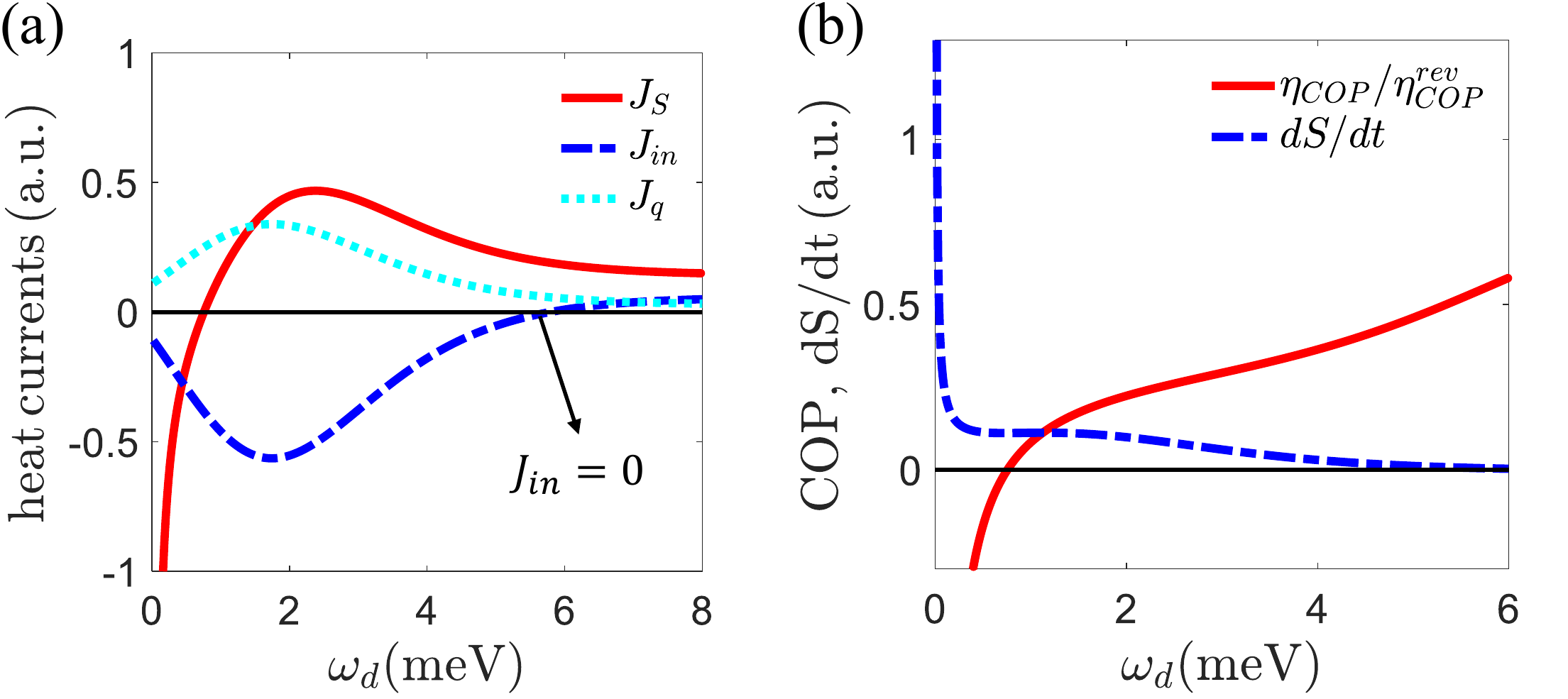}
\caption{(a) Thermal currents and, (b) COP ratio $\eta_{COP}/\eta^{rev}_{COP}$ and entropy production $dS/dt$ as a function of the QD energy $\omega_d$. Other parameters: $\mu=0$, $k_BT_D=1\,\rm{meV}$, $k_BT_H=1.5\,\rm{meV}$, $k_BT_S=0.9\,\rm{meV}$, $E_1=E_4=1\,\rm{meV}$ and $T_C=1/(2/T_D-1/T_H)$.}
\label{fig:cu-wd}
\end{center}
\end{figure}

\begin{figure}[htb]
\begin{center}
\centering \includegraphics[width=8.5cm]{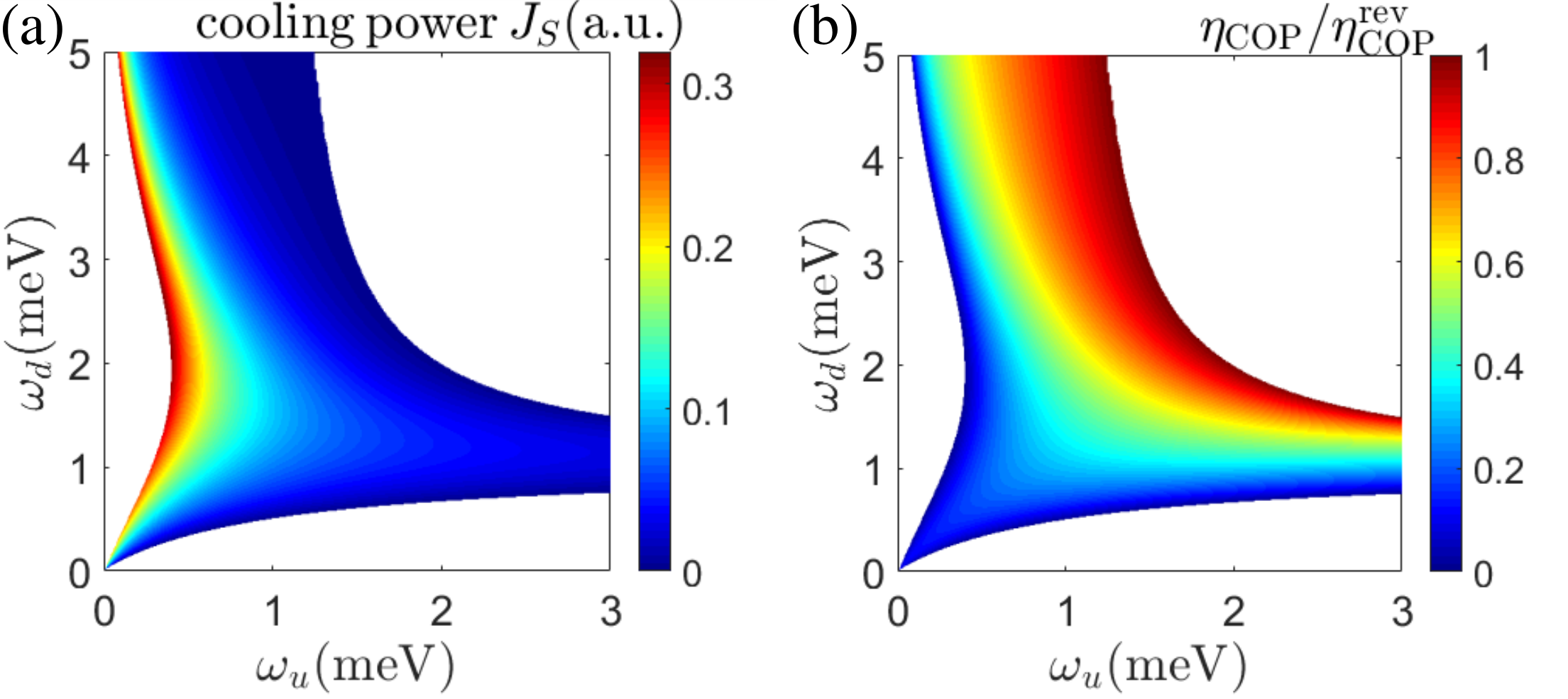}
\caption{(a) cooling power and (b) COP of the cooling by transverse current effect as functions the two energies $\omega_u$ and $\omega_d$ for $J_{in}=0$. Other parameters: $\mu=0$, $k_BT_D=1\,\rm{meV}$, $k_BT_H=2.5\,\rm{meV}$, $E_1=E_4=1\,\rm{meV}$ and $T_C=1/(2/T_D-1/T_H)$.}\label{fig:demon}
\label{fig:PE-ud}
\end{center}
\end{figure}

We first show that the cooling by transverse current effect surves even when $J_{in} = 0$ (i.e., the total heat current injected into the quantum system vanishes). In Fig.~\ref{fig:cu-wd}(a), we show the three thermal currents, $J_S$, $J_q$, and $J_{in}$, as functions of $\omega_d$ with other parameters fixed and given in the caption of the figure. We focus on the conditions where $A_S < 0$, $A_{in} = 0$, and $A_q > 0$. It is seen from Fig.~\ref{fig:cu-wd}(a) that there is indeed a special point, $\omega_d\simeq 5.7$~meV, where the cooling by transverse current effect surves at $T_S<T_D$ and $J_{in} = 0$. We also notice from the figure that the sign change of $J_{in}$ does not affect the cooling by transverse current effect. The COP $\eta_{COP}$ and the cooling power $J_S$ remains positive for $\omega_d> 5.7$~meV, regardless of the heat current $J_{in}$. However, only the case with $J_{in} = 0$ represents the nonequilibrium demon.

Figure~\ref{fig:demon} presents the cooling power $J_S$ and COP as functions of the QD energies $\omega_u$ and $\omega_d$ at the condition with $J_{in}=0$. The white areas represent the parameter regions where the cooling by transverse current effect cannot be achieved at the condition with $J_{in}=0$. As shown in Fig.~\ref{fig:demon}(a), the high cooling powers can be achieved in the region with $\omega_u< 1$~meV. However, in such a region, the cooling efficiency is small. In fact, the cooling efficiency is large when $\omega_d> 1$~meV, reflecting the power-efficiency trade-off in the nonequilibrium demon device~\cite{JiangPRE,PED}.

\section{conclusion and discussions}
We have demonstrated a mode of cooling, the cooling by transverse current effect, using a four-terminal (i.e., the source, the dain, and two thermal baths) QDs thermoelectric device. Such an effect describes cooling of the source driven by the heat exchange between the two thermal baths, rather than the total heat injected into the quantum system~\cite{WhitneyPhysE}. A simple realization is to use a device with four QDs which simultaneously breaks the particle-hole symmetry, the left-right and up-down inversion symmetries. The cooling power and efficiency demonstrate trade-off effect, e.g., high efficiency comes with low cooling power, consistent with the common intuitions.

In addition, a transverse thermoelectric effect, i.e., generating electric power using the temperature difference between the two heat baths is studied. In the transverse thermoelectric effect, electric and heat transport are spatially separated. Such a thermoelectric effect has the advantage of manipulation of heat and charge transport in different spatial dimensions. Such spatial separation enables disentangling the correlations between electric conduction and phonon heat conduction, so that they can be engineered independently to yield high figure of merit and output power. The current-affinity relations, linear-transport properties, and the transverse thermoelectric figure of merit are studied.

We also show that the four-terminal quantum-dot system can realize a type of Maxwell demon where the demon exploits nonequilibirium effects to drag heat from a low-temperature reservoir (i.e., the source) to a high-temeperature reservoir (i.e., the drain), even when there is no net energy and charge exchange between the demon and the system. Our findings demonstrate that inelastic transport can bring about phenomena that have not been found in previous studies based on elastic transport processes.

\section{Acknowledgment}
We acknowledge support from the National Natural Science Foundation of China (NSFC Grant No. 11675116), the Jiangsu distinguished professor funding and a Project Funded by the Priority Academic Program Development of Jiangsu Higher Education Institutions (PAPD). 

\bibliography{Ref-demon}

\end{document}